\documentclass[amsmath,amssymb,aps,prb,reprint,groupedaddress]{revtex4-2}
\usepackage{graphicx}
\usepackage{dcolumn}
\usepackage{bm}
\usepackage{xcolor}

\begin{document}

\title{Angle-dependent electron confinement in graphene moir\'e superlattices}
\author{Francisco S\'anchez-Ochoa} 
\author{Andr\'es R. Botello-M\'endez}
\author{Cecilia Noguez}
\email[Corresponding author: ]{cecilia@fisica.unam.mx}
\affiliation{Instituto de F\'{i}sica, Universidad Nacional Aut\'{o}noma de
   M\'{e}xico, Apartado Postal 20-364, Cd. de M\'{e}xico C.P. 01000,
   Mexico}

\date{\today}

\begin{abstract}
In graphene moir\'e superlattices, electronic interactions between layers are mostly hidden as band structures get crowded because of folding, making their interpretation cumbersome. Here, the evolution of the electronic band structure as a function of the interlayer rotation angle is studied using Density Functional Theory followed by unfolding bands and then comparing them to their corresponding individual components. We  observe interactions at regions not theoretically elucidated so far, where for small interlayer angles, gaps turn into discrete-like states that are evenly-spaced in energy. We find that $V_{pp\sigma}$ attractive interactions between out-of-plane orbitals from different layers are responsible for the discretization. Furthermore, when the interlayer angle becomes small, these discrete evenly-spaced states have energy differences comparable to graphene phonons. Thus, they might be relevant to explain electron-phonon assisted effects, which have been experimentally observed in graphene moir\'e superlattices.
\end{abstract}

\maketitle

\section{Introduction}
The vertical stacking of any two-dimensional (2D) layered materials, such as graphene (G), transition metal dichalcogenides, hexagonal boron nitride, or layered oxides,\cite{vdw1} has led to a new research field with exciting phenomena and potential applications.\cite{vdw2} These atomic 2D films are known as van der Waals heterostructures, following the weak forces nature that maintains the layers together. This weak interaction suggests that the combined 2D systems' electronic properties could be a simple superposition.\cite{vdw3} However, recent studies that focused on atomic bilayers have shown that the physical properties can drastically modify because of the interlayer interaction, which additionally can be modulated with the relative angle between layers.\cite{cao2018unconventional,Cao:2018wf,yankowitz2019tuning,georgiou2013vertical,britnell2012field,hidalgo2019tuning,vdw20,vdw21,vdw22,vdw42,brihuega2012unraveling,vdw6}

A lattice mismatch occurs when two G layers are vertically stacked with a different crystallographic orientation between them. When a periodicity appears upon stacking, a superlattice arrangement arises with a characteristic moir\'e pattern.\cite{PhysRevB.84.035440} The interference effect between the two layers with slightly different orientations, i.e.,  a minimal interlayer angle, generates moir\'e patterns with long-wavelengths. Thus, a graphene moir\'e superlattice is a periodic 2D material composed of twisted-bilayer G (TBLG). Here, new physical effects arise from the interlayer potential modulation that depends on the relative orientation. Moir\'e patterns are composed of local combinations of vertical atomic arrangements AA and AB (BA), these latter have zero interlayer angles, and sequences between them. The arrangements fluctuations modulate the interlayer potential and, consequently, the electronic band structure of TBLG. \cite{hunt2013massive,li2010observation,vdw7} In AA arrangements, the atoms in both layers are precisely aligned, while in AB (BA) arrangements, the atoms of one layer lie directly on the center of a hexagon in the lower (upper) graphene sheet. The AB order is also known as Bernal stacking, and it is more stable than the
AA-stacked bilayer. \cite{vdw5} Density functional theory (DFT) and semi-empirical tight-binding studies have shown that AA and AB arrangements show different electronic behavior near the Fermi level $(E_F)$, while at lower energies their bands are quite similar.\cite{vdw5}

Experimental results obtained using different characterization techniques, such as scanning tunneling microscopy (STM), Raman, and angle-resolved photoelectron spectroscopy (ARPES), among others, have shown that TBLGs exhibit band structures different from the isolated layer. \cite{yang2018visualizing,vdw4b,brihuega2012unraveling,PhysRevLett.109.186807,lisi2020direct,Havener2012,HeRui2013} It suggests an interlayer coupling that modifies the electronic band structure near $E_F$, leading to distinct novel electronic and optical properties compared with the isolated G layer. \cite{vdw3,vdw4,moon2013optical} Some exciting results include van Hove singularities (VHS), mini gaps in the far-infrared region, a decrease in charge-carriers velocity near the Dirac point, the so-called moir\'e bands, localization of Dirac electrons, as well as stacking-dependent optical and conduction properties, together with an angle-controlled optical activity, among others.
\cite{vdw3,vdw4,moon2013optical,cao2018unconventional,Cao:2018wf,yankowitz2019tuning,georgiou2013vertical,britnell2012field,hidalgo2019tuning,vdw20,vdw21,vdw22,vdw42,brihuega2012unraveling,vdw6,vdw5,PhysRevB.84.035440,yang2018visualizing,vdw4b,PhysRevLett.109.186807,lisi2020direct,Havener2012,HeRui2013}

In this work, the evolution of the electronic band structure as a function of the interlayer rotation angle ($\theta$) in graphene moir\'e superlattices (GMSs) is studied. We employed a methodology that allows us to elucidate the physical effects of the mutual interactions between 2D layers by unfolding the electronic bands of systems with double-periodicity.\cite{vdw11} In particular, the combination of total energy DFT calculations followed by the unfolding approach permits us to obtain effective electronic bands (EEBs) projected onto the isolated G primitive cell. Then, EEB allows us to give a direct physical interpretation of the electronic changes that a G layer suffers upon the presence of a second sheet. Applying the above methodology, we identify many energy band gaps at about 2~eV, above and below $ E_F $, close to $\mathbf{M}$, and around a well-defined energy window. These states exhibit a discrete-ladder-like behavior at small interlayer angles induced by the electronic degeneration breaking, caused by  $V_{pp\sigma}$ interactions  between $\pi-$band electrons from each layer. At lower energies and around $\mathbf{\Gamma}$, a band splitting is found by inspecting the EEB. At different energies and reciprocal $\mathbf{k}-$points, all these electronic effects  are associated with angle-modulated interactions between the out-of-plane orbitals of both G layers that form the $\pi-$bands. Additionally, we can easily reproduce and identify the Dirac cone interactions between layers, which are recognized as van Hove singularities that move toward $ E_F $ as $\theta$ decreases. This effect has been observed experimentally and theoretically.\cite{brihuega2012unraveling,moon2013optical,PhysRevLett.109.186807}  The results discussed here provide a precise physical analysis of the electronic band structure evolution of TBLG as a function of the relative angle between layers, where discrete evenly-spaced states and band splitting are observed, besides the well-known Dirac cone interactions and variation of the charge carriers group velocity.

\section{Methods}
\subsection{TBLG structures}\label{TBGL}

Commensurate GMSs or TBLG are labeled with two integer indices, $(m,n)$ associated with the G primitive cell lattice vectors on each layer. The $(m,n)$ indices are used to construct TBLG systems, as well as the matrix transformation, $\mathbf{P}$, used in the unfolding method.\cite{vdw11} The structural parameters such as $\theta~(\circ)$, the interlayer rotation angle; $N$, the total number of carbon atoms in GMS; $a_{\textsc{sc}}$, superlattice cell parameter; and $R$, reconstruction factor, are calculated as in Ref.~\citenum{vdw24} by using the following expressions:
\begin{equation}
\label{eqn9}
\theta = \cos^{-1}\left[ \frac{m^2+4mn+n^2}{2\left(m^2+mn+n^2\right)} \right]*\left( \frac{180}{\pi}
\right) \, ,
\end{equation}
\begin{equation}
\label{eqn10}
N = 4 \left( m^{2}+n^{2}+mn \right) \, ,
\end{equation}
\begin{equation}
\label{eqn8}
a_{\textsc{sc}} = a_{PC} \sqrt{\vert \det\mathbf{P} \vert} \, ,
\end{equation}
\begin{equation}
\label{eqn11}
R = \sqrt{\vert \det\mathbf{P} \vert} \times \sqrt{\vert \det\mathbf{P} \vert} \, ,
\end{equation}
respectively. Here, $m$ and $n$ are integer numbers that characterized commensurate GMS, where $m \geq n$. Also, $a_{PC}$ = 2.44~\r{A} is the lattice constant of the optimized G primitive cell, $\det$ means the determinant of a matrix, and $\mathbf{P}$ is the matrix used in the linear transformation between
primitive cell and supercell (\textsc{sc}) lattice vectors, according to Ref.~\citenum{vdw11}. Here, we study TBLG or commensurate GMS with $(m,n)$ listed in Table~\ref{table1}.

\begin{table}
  \caption{Structural parameters of GMS defined with $(m,n)$. $\theta$, relative angle; $N$ number
    of carbon atoms in the supercell with a lattice constant $a_{\textsc{sc}}$, and reconstruction
    factor $R$. After optimization, $d$: average distance between layers, and $\Delta$: average
    atomic corrugation.}
  \label{table1}
  \begin{ruledtabular}
  \begin{tabular}{crrrcrr}
    $(m, n)$ & $\theta$~$(^{\circ})$ & $N$ & $a_{\textsc{sc}}$~(\r{A}) & $R$ & $d$~(\r{A}) & $\Delta$~(m\r{A})\\
    \hline
    (2,1)   & 21.78 & 28   & 6.45  & $\sqrt{7}\times\sqrt{7}$               & 3.28 & 1.89 \\
    (4,2)   & 21.78 & 112  & 12.91 & $2\sqrt{7}\times2\sqrt{7}$             & 3.28 & 2.44 \\
    (5,3)   & 16.42 & 196  & 17.07 & $7\times7$                             & 3.28 & 16.36 \\
    (3,2)   & 13.17 & 76   & 10.63 & $\sqrt{19}\times\sqrt{19}$             & 3.26 & 23.89 \\
    (6,4)   & 13.17 & 304  & 21.27 & $2\sqrt{19}\times2\sqrt{19}$           & 3.29 & 26.86 \\
    (7,5)   & 10.99 & 436  & 25.47 & $\sqrt{109}\times\sqrt{109}$           & 3.31 & 37.33 \\
    (4,3)   & 9.43  & 148  & 14.84 & $\sqrt{37}\times\sqrt{37}$             & 3.29 & 40.18 \\
    (5,4)   & 7.34  & 244  & 19.05 & $\sqrt{61}\times\sqrt{61}$             & 3.30 & 61.01 \\
    (6,5)   & 6.01  & 364  & 23.27 & $\sqrt{7\cdot13}\times\sqrt{7\cdot13}$ & 3.30 & 69.12 \\
    (7,6)   & 5.09  & 508  & 27.49 & $\sqrt{127}\times\sqrt{127}$           & 3.30 & 69.82 \\
    (9,8)   & 3.89  & 868  & 35.94 & $\sqrt{7\cdot31}\times\sqrt{7\cdot31}$ & 3.31 & 72.19 \\
    (11,10) & 3.15  & 1324 & 44.39 & $\sqrt{331}\times\sqrt{331}$           & 3.29 & 74    \\
    (13,12) & 2.64  & 1876 & 52.84 & $\sqrt{7\cdot67}\times\sqrt{7\cdot67}$ & 3.32 & 72.17
  \end{tabular}
 \end{ruledtabular}
\end{table}

\subsection{DFT calculations}\label{DFT}

Total energy DFT calculations were performed using the SIESTA code. \cite{ordejon1996self,soler2002siesta} The exchange-correlation energy was described by a van der Waals (VDW)\cite{roman2009efficient}
functional using the Klime{\v{s}}, Bowler and Michaelides parametrization.\cite{klimevs2009chemical} The electron-ion interactions are treated with norm-conserving pseudopotentials.\cite{troullier1991efficient} A linear combination of pseudoatomic orbitals (LCAO) is employed to expand the valence electronic states, with an optimized double-$\zeta$ polarized (DZP) basis set. An energy mesh-cutoff of 500~Ry is applied to sample the electronic density in the real space. The electronic self-consistency was converged to $10^{-5}$ value. The Monkhorst--Pack scheme\cite{monkhorst1976special} is used to sample the Brillouin zone with an optimal $21\times21 \times1$ grid for the smallest \textsc{sc}. The atomic optimization was carried out with the conjugate gradient algorithm with a maximum value in the interatomic forces of 0.01~eV/\r{A}. Energy cutoff and \textit{k}-grid convergence tests are performed on the systems leading to the values above as the optimal ones. To simulate a GMS, we employ the \textsc{sc} method with a vacuum space of 25~\r{A} between the adjacent G monolayers to avoid spurious interactions. The interlayer angle between G monolayers takes a reference to the zigzag directions of each sheet, as reported in the Ref.~\citenum{vdw24}. Since both G sheets have hexagonal symmetry, then the maximum limit of the rotation angle is $\theta=30^{\circ}$. Visualization of atomic models and isosurfaces are performed using VESTA.\cite{momma2011vesta}

\subsection{Unfolding method}
We employ a general unfolding method for the electronic bands of systems with double-periodicity that let elucidate the physical effects of the mutual interactions between systems. The method is based on DFT with a linear combination of atomic orbitals as a basis set and considers two symmetry operations of the primitive cell: a standard expansion, and a single rotation. As a result, the unfolding method enables studying the electronic properties of vertically stacked homo or heterostructures. The details of the unfolding approach are discussed in Ref.~\citenum{vdw11}. Briefly, the unfolding method follows the next steps: (i) The electronic band structure is calculated in the BZ of the \textsc{sc}, and then, it is  projected into the reciprocal space of the primitive cell, i.e., eigenvalues and eigenfunctions, $E(\mathbf{k}), \psi_I(\mathbf{k},\mathbf{r})$, where $I$ labels the different bands. (ii) Then, we identify the equivalent atoms between the extended and primitive cells, which allows us to determine the corresponding eigenfunctions, which are orthonormalized. (iii) Next, we project the eigenfunctions extended cell onto the primitive cell ones using the transformation matrix, $\mathbf{P}$, and then, calculating the spectral weights, $W(\mathbf{k},I)$. (iv) Finally, the effective electronic band (EEB) is obtained along a path that contains the high-symmetry points of the BZ of the primitive cell. The spectral weights, $W(\mathbf{k},I)$, determine how much of the pristine G electron state is preserved upon the presence of a second layer. Thus, $W(\mathbf{k},I)=1$ means that the G electronic state is fully recovered, while $W(\mathbf{k},I)=0$ indicates is absent. Meanwhile, $W(\mathbf{k},I)<1$ suggests an alteration of the electronic state because of the interlayer interaction. In this way, EEB compares directly to the band structure of pristine G. This latter and its density of states (DOS) are shown for reference in Fig.~\ref{S1} in Appendix \ref{graph} for completeness. This method opens the possibility to directly compare with experiments (for example, ARPES spectra), which otherwise would be impossible to distinguish.

\section{Results and Discussion}

\subsection{Structure of TBLG as a function $\theta$}

\begin{figure}[t]
  \includegraphics[width=0.5\textwidth]{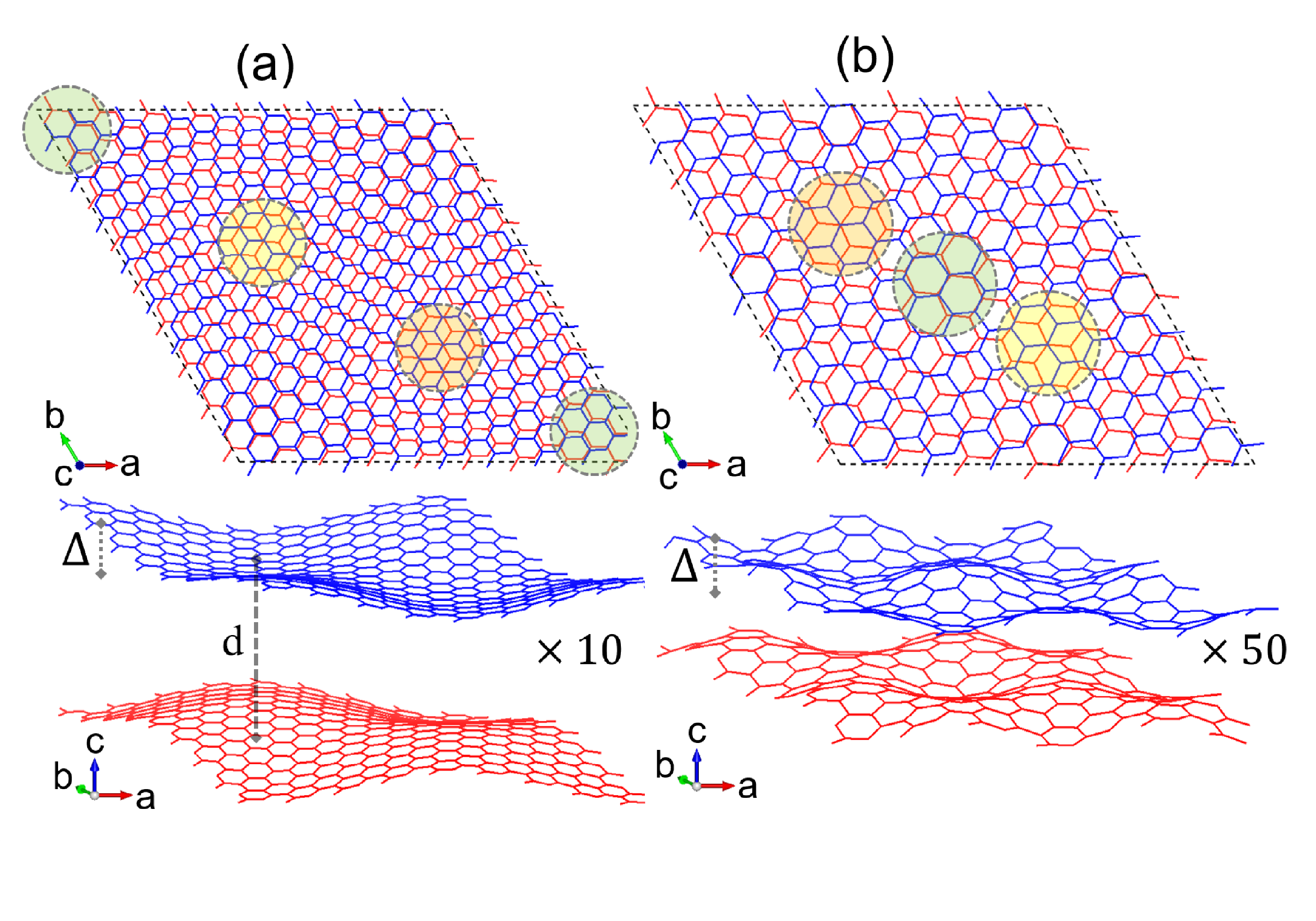}
  \caption{Optimized TBLG models: (a) $(9,8)$ and (b) $(7,5)$. Red and blue lattices represent bottom and top G layer, respectively. Upper panels show the top views, where AA$-$, AB$-$, and BA$-$ domains are
    indicated with green, yellow, and orange circles, respectively. Bottom panels show isometric
    views scaled up (a) $10$  and (b) $50$ times. $\Delta$ is the atomic corrugation and $d$ is the average difference of the vertical coordinate on each G layer.}
  \label{Fig1}
\end{figure}

Let us start discussing the structural parameters of TBLGs upon full optimization using \textsc{siesta}.\cite{ordejon1996self,soler2002siesta} We first consider an AA$-$stacking pattern with $\theta=0^\circ$, followed by a rotation between layers. While, TBLG superlattices exhibit AA$-$, AB$-$, and BA$-$stacking patterns in different regions. These three patterns are indicated with green, yellow and orange circles, respectively in the upper panels of Fig.~\ref{Fig1}. Here, top views for (a) $(9,8)$, and (b) $(7,5)$ TBLGs are shown. Their corresponding isometric views are in bottom panels, where they are scaled up by 10 and 50 times, respectively, to make visible the atomic corrugation after optimization. From the isometric views, sinusoidal patterns in the atomic corrugation $(\Delta)$ are found, with maxima separations in AA$-$stacking domains, and minima in AB$-$ or BA$-$ ones. This sinusoidal behavior of $\Delta$ has the same periodicity as the moir\'e supercell and agrees with previous reports.\cite{uchida2014atomic} The number of AA, AB, and BA domains increases as $\theta \to 0^\circ$. But when $\theta=0^\circ$ only the AA$-$stacking is present and a null corrugation is observed. The quantified average distances, $d$, between monolayers present variations smaller than $2\%$ as a function of the angle. However, $\Delta$ increases nearly two orders of magnitude as $\theta$ goes to zero, being smaller than 100~m\r{A} in all cases. Notice that TBLG with different $(m, n)$ may have the same relative angle but different $N$, $a_{\textsc{sc}}$, and $R$, where all these parameters are related by an integer factor. For example, TBLGs $(2, 1)$ and $(4, 2)$ in Table~\ref{table1} have same $\theta=21.78^\circ$, while $N_{(4,2)}=4 N_{(2,1)}$, etc. However, independently of $N$, TBLGs $(2, 1)$ and $(4, 2)$ have the largest $\theta$ but the smallest $\Delta$ because they have exactly the same moir\'e pattern. In this case, the in-plane relaxation is about 0.001~\AA, which is negligible.

\subsection{Effective electronic bands (EEBs) in TBLG}

\begin{figure}[t]
  \includegraphics[width=0.5\textwidth]{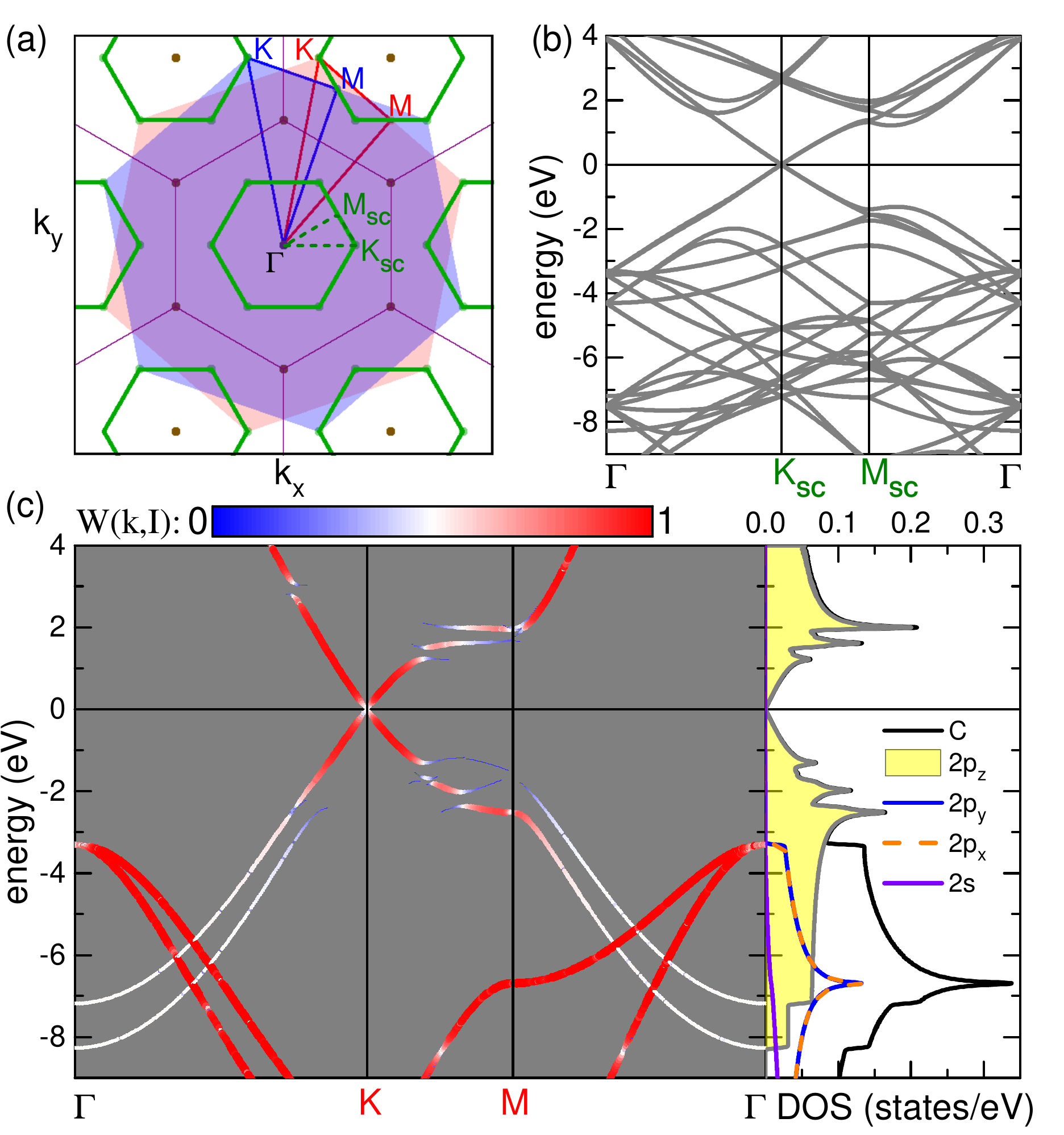}
  \caption{TBLG $(2,1)$ with $\theta = 21.78^{\circ}$. (a) BZ of \textsc{sc} in green-solid lines, of both G primitive cells are in red and blue hexagons. Their $\mathbf{K}$ and $\mathbf{M}$ high-symmetry points are also in green, red, and blue colors. $\mathbf{\Gamma}$ coincides for all BZs. (b) Calculated electronic bands of \textsc{sc} (green-dashed lines in (a)). (c) Left panel: unfolded electronic bands or EEB of \textsc{sc} projected onto one G primitive cell, where $W(k,I)$ is proportional to the size and color of the dots (color bar on the top). Right panel: DOS per atom of \textsc{sc}, total (black line) and partial contributions from orbitals. }
  \label{Fig2}
\end{figure}

To discuss the electronic properties, we illustrate the unfolding method used here for TBLG  $(2,1)$ with $\theta=21.78^{\circ}$. Notice that TBLGs with  same twist angle but different $(m,n)$ have ideltical electronic properties, as expected. The Brillouin Zones (BZs) corresponding to \textsc{sc}, as well as those of top and bottom G primitive cells are shown in Fig.~\ref{Fig2}(a) in green, red and blue hexagons. For each BZ their high-symmetry points, $\mathbf{K}$ and $\mathbf{M}$, are indicated with the corresponding color. While $\mathbf{\Gamma}$ is the same in all BZs. Fig.~\ref{Fig2}(b) shows the electronic bands of the \textsc{sc} along the path in dashed-green lines in Fig.~\ref{Fig2}(a), $\mathbf{\Gamma} \to \mathbf{K}_{\text{sc}} \to \mathbf{M}_{\text{sc}} \to \mathbf{\Gamma}$. Electronic states from both G layers are folded into the reciprocal space of \textsc{sc}.\cite{ashcroft1976solid} We can identify the linear dispersion of Dirac cones close to $E_F$ at $\mathbf{K}_{\text{sc}}$, where $E_F$ was set $0$~eV in all graphs for simplicity. However, the change of the electronic structure is hidden by the use of the \textsc{sc}. Consequently, it is hard to go deeper in our analysis from Fig.~\ref{Fig2}(b), because of the large number of bands due to folding.

To overcome the problem, we employ the unfolding approach described above. The left-hand side of Fig.~\ref{Fig2}(c) shows EEB, where the size and color of the dots represent $W(k,I)$. The bands forming the Dirac cones show discontinuities with $W(k,I)<1$, in both valence and conduction bands. Following these bands at deeper energies, a split into two bands around $-8$~eV toward $\mathbf{\Gamma}$ is found, with $W(k,I)\simeq0.5$ for each one and an energy difference of $\sim$1~eV. In contrast, the rest of the bands are practically unaffected, with $W(k,I)\simeq1$. The effect of interlayer interaction on $\sigma$-bands is expected to be small because of the orthogonal character of these bands to the $\pi$-bands involved in the interaction. Using the information from the total and partial DOS in the right-hand panel of Fig.~\ref{Fig2}(c), we found that the main alterations in EEB are along the bands associated with $2p_{z}$ or out-of-plane orbitals, known as $\pi-$bands. On the other hand, the unchanged bands result from $2s$, $2p_{x}$, and $2p_{y}$ or in-plane orbitals combinations, and are knows as $\sigma-$bands. Between $(-3.0,-1.5)$~eV and $(+1.2,+2.5)$~eV, significant differences in the $\pi-$bands reveal peaks in DOS above and below $E_F$, corresponding to those energies where $\pi-$bands are broken and given origin to van Hove singularities (VHS). At lower energies, a step-function behavior is observed in DOS, associated with the band splitting. The half drop of DOS at $-7.1$~eV is consistent with $W(k,I)=0.5$, the spectral weight results. Note that from the electronic bands of the \textsc{sc} in Fig.~\ref{Fig2}(b), it is impossible to identify the origin of most VHS and DOS step-function below $-7$~eV, besides few features close to $E_F$. 

In general, we find that $\pi-$bands are perturbed; meanwhile, $\sigma-$bands ones remain unaffected. The alteration of the $\pi-$bands is due to the strong electronic interaction between $2p_{z}$ or out-of-plane orbitals belonging to different layers. The different numbers and positions of the AA$-$, AB$-$, and BA$-$stacking domains found at different angles modulate the out-of-plane interactions. To demonstrate it, in Fig.~\ref{S2} in Appendix \ref{graph}, we show the EEB of \textsc{sc} TBLG $(2,1)$ with an interlayer distance more than twice the equilibrium distance and its corresponding DOS. Here, the electronic bands of pristine G are reproduced. Therefore, we can conclude that the significant alterations in the $\pi-$bands observed in EEB and its DOS at the equilibrium interlayer distance are caused by the strong electronic interaction between $2p_{z}$ orbitals and not because of Bragg diffractions inherent to the \textsc{sc} folding.

\subsection{EEBs as a function of $\theta$}

\begin{figure}[t]
  \includegraphics[width=0.47\textwidth]{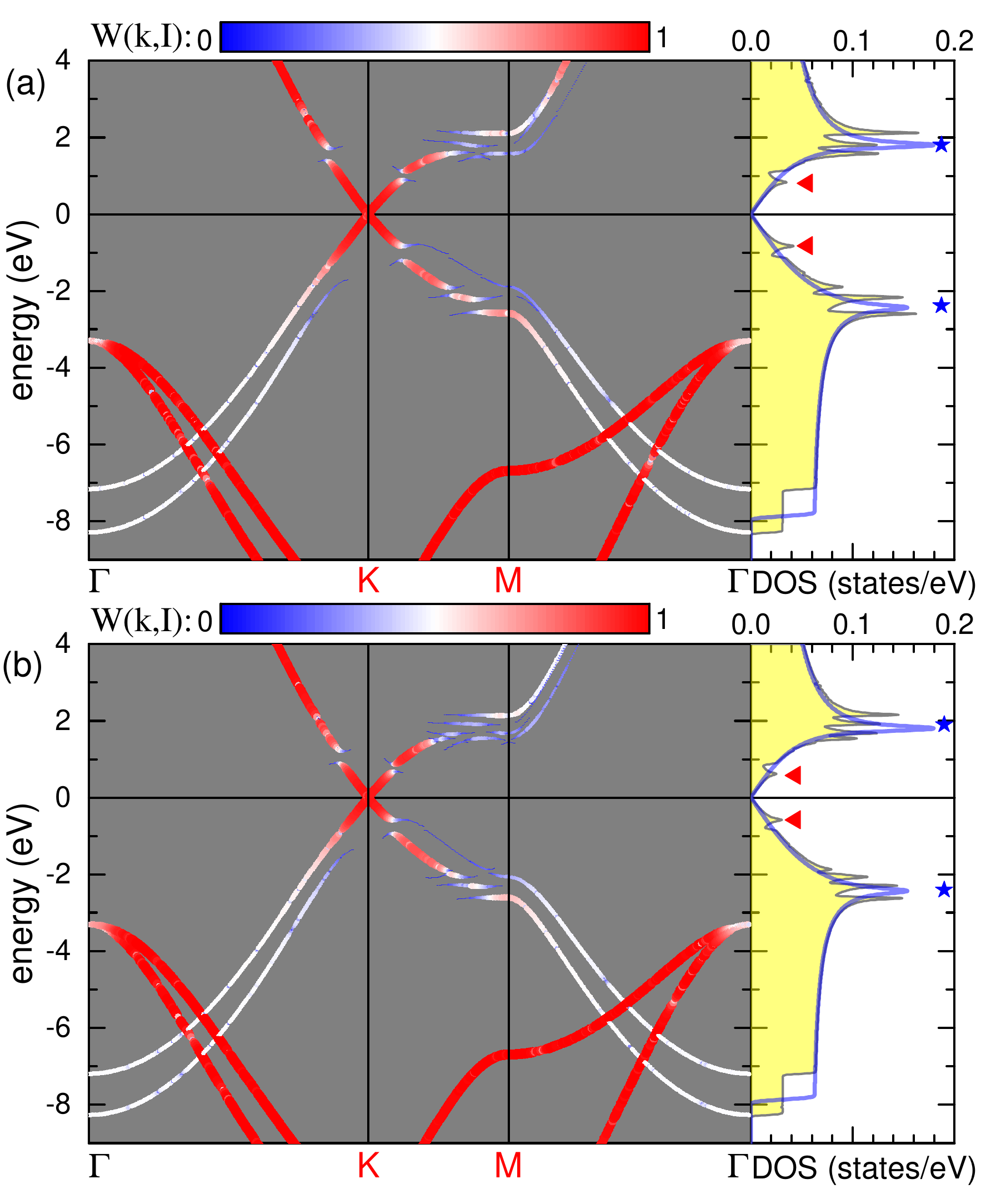}
  \caption{EEB of TBLGs with angles (a) $\theta = 13.17^{\circ}$ and $(3,2)$, and (b) $\theta = 9.43^{\circ}$ and $(4,3)$. Partial DOS per atoms with contributions of $2p_z$ is included (yellow), and also of pristine G (blue line). Red triangles mark VHS close to $\mathbf{K}$ and $E_F$, while blue stars indicate VHS from gaps close to $\mathbf{M}$.}
  \label{Fig3}
\end{figure}

To analyze the electronic properties angle dependence, we consider two additionally TBLG with $\theta = 13.17^{\circ}$ and $\theta = 9.43^{\circ}$. Applying the above procedure, we calculate their EEB and DOS, shown in Fig.~\ref{Fig3}(a) and (b), respectively.  
Again, $\sigma-$bands remain unperturbed, the $\pi-$band splitting is observed around $-8$~eV at the $\Gamma$ point, showing quite similar features as for $\theta = 21.78^{\circ}$, and also $\pi-$bands exhibit clear modifications that depend on the given angle. For instance, we find quasi-flat bands at about $(-3.0,-1.5)$~eV and $(+1.2,+2.5)$~eV in the $\mathbf{K}-\mathbf{M}$ segment that give rise to $\pi-$band discontinuities and VHS in DOS. It is important to notice that the energies, $k-$points, and the number of discontinuities or gaps depend on the relative interlayer angle. For example, the gaps at low energies above and below $E_F$, and marked with a red triangle in the DOS plots in Fig.~\ref{Fig3}, move simultaneously toward $\mathbf{K}$ and $E_F$ as $\theta$ turns small. This latter result is in excellent agreement with previous experimental and theoretical reports, which identified these VHS as the consequence of the interactions between Dirac cones in each layer. \cite{cao2018unconventional,moon2013optical,brihuega2012unraveling,li2010observation} 
In Appendix \ref{dirac}, we show our main findings for these gaps close to $\mathbf{K}$ and $E_F$ as a function of $\theta$, which compare very well with previous experimental and theoretical results.

On the other hand, the  mini-gaps around $ \pm 2.5$~eV behave differently as a function of $\theta$. They are characterized by three and four VHS for $\theta = 13.17^{\circ}$ and $\theta = 9.43^{\circ}$, respectively. They are closer to $\mathbf{M}$ instead of $\mathbf{K}$, and remain around the same energies, almost independent on the relative angle. However, as a function of $\theta$ they undergo major changes at energies around $-3.0$ and $+3.0$~eV and in $\mathbf{K}-\mathbf{M}$. The many gaps that appear around $-2.5$~eV for occupied states, and $+2.0$~eV for empty states close to $\mathbf{M}$ coming from $\mathbf{K}$, behave differently than those that move simultaneously toward $\mathbf{K}$ and $E_F$ as $\theta$ turns small. In pristine G (Fig.~\ref{S1} in Appendix \ref{graph}), we see that the $\pi-$band shows saddle points at about $-2.2$~eV and $+1.8$~eV in $\mathbf{M}$ and consequently VHS in DOS. However, in the presence of a second G layer, these bands split into many discrete-like states. Furthermore, as the relative angle diminishes, more localized states appear, spreading over an energy window not larger than 1.0~eV.  These states show an almost constant energy separation between successive VHS peaks, as seen in Fig.~\ref{S3}(a) in Appendix \ref{dirac} between $-2.8$ and $-2.0$~eV for occupied states, and between $+1.5$ and $+2.3$~eV for empty states. 

\subsection{Origin of discrete electronic states close $\mathbf{M}$}

To elucidate the origin of these states, we additionally perform semi-empirical tight-binding calculations. Here, $V_{pp\sigma}$ (attractive) and $V_{pp\pi}$ (repulsive) overlapping parameters of $2p_z$ orbitals in both G layers are considered. By switching on and off these parameters and looking in-plane and out-of-plane interactions, we find that $V_{pp\sigma}$ attractive interactions between out-of-plane orbitals from different layers are responsible of the discretization. The equally-spaced energy distribution between consecutive states suggests an electronic resonant behavior between layers, resembling quantum dipole oscillators observed in G quantum dots. \cite{doi:10.1021/acs.chemrev.6b00446} In Appendix \ref{tb}, we show a brief summary of the tight-binding model we employed. The interlayer interactions are observed by switching on and off in-plane and out-of-plane parameters for the electronic bands, as shown in Fig.~\ref{S4}.

\begin{figure}[t]
  \includegraphics[width=0.50\textwidth]{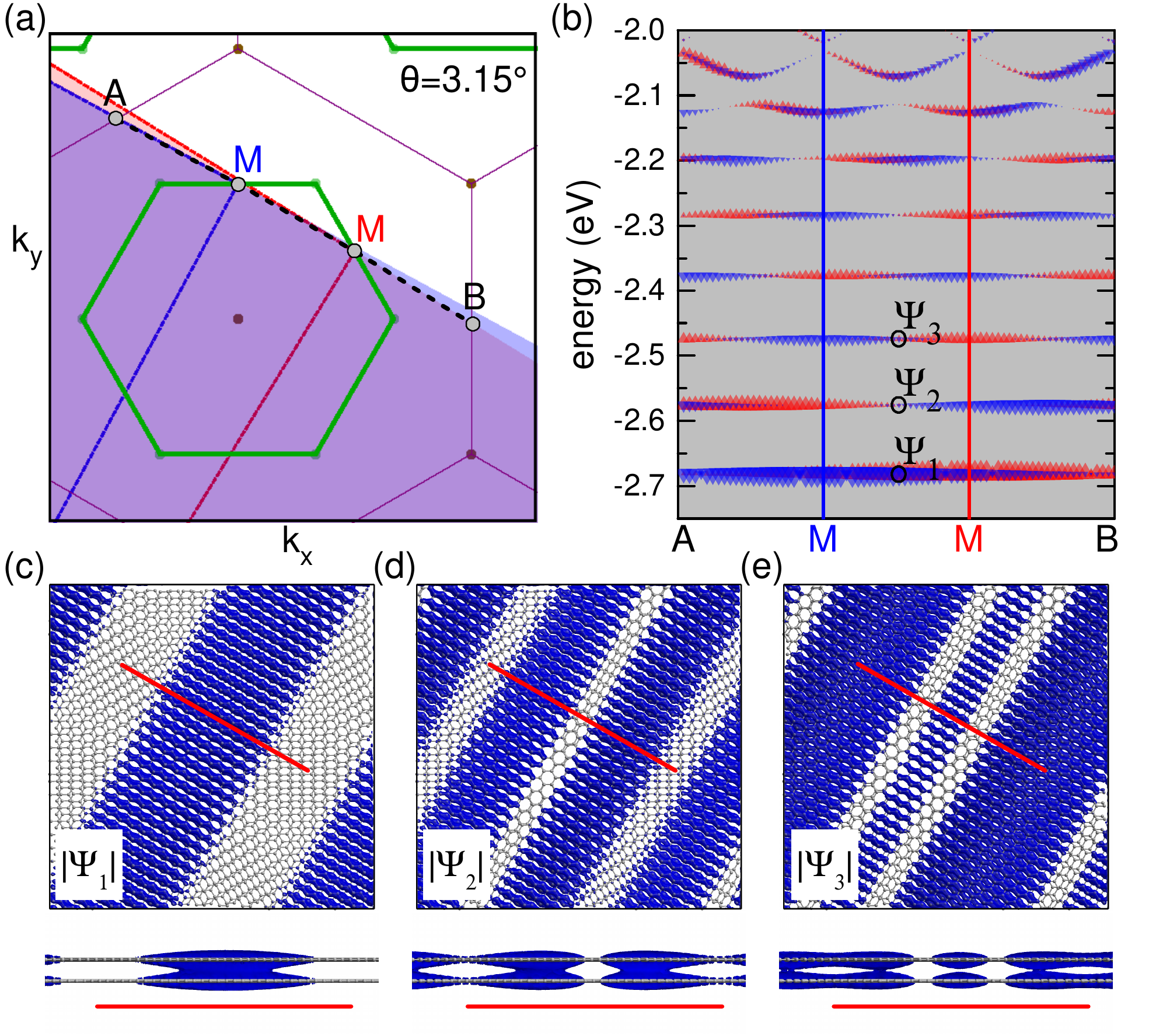}
  \caption{(a) \textsc{sc} reciprocal space showing the $\mathbf{M}$ points belonging to the upper (blue) and bottom (red) G layers. (b) EEB along $\mathbf{A} \to \mathbf{M}_{\text{blue}} \to \mathbf{M}_{\text{red}} \to \mathbf{B}$ for TBLG with $\theta=3.15^\circ$, projected at the upper (blue) and bottom (red) G layers. $W(k,I)$ is proportional to triangles size. (c), (d) and (e) show top views of the real space of about $60$~\AA, side, along with the corresponding $|\Psi|$ evaluated at the middle of $ \mathbf{M}_{\text{blue}} \to \mathbf{M}_{\text{red}}$ for the electronic states with lower energy, as indicated in (b). Below the squares, we show the side views of the  the corresponding $|\Psi|$, which are projected along the red line of about 40~\AA, show in the top views. Here, isosurfaces are calculated with $0.01$~\AA$^{-2}$.}
  \label{Fig4}
\end{figure}

To observe these equally-spaced bands, we analyze the projected states along the trajectory between both $\mathbf{M}$ points that belong to the upper and bottom G sheets. This is shown in Fig.~\ref{Fig4}(a), where red and blue colors are used to identify in which G layer is situated the $\mathbf{M}$ point. As $\theta$ decreases, the distance between both $\mathbf{M}$ points also does. Here, we show the projected \textsc{sc} states in both upper and bottom G sheets when the interlayer angle is just $\theta=3.15^\circ$. The projection is performed along the $\mathbf{A} \to \mathbf{M}_{\text{blue}} \to \mathbf{M}_{\text{red}} \to \mathbf{B}$, which corresponds to the zig-zag direction in the real space and $\mathbf{A}$ and $\mathbf{B}$ are shown in Fig.~\ref{Fig4}(a). Now, in Fig.~\ref{Fig4}(b), we observe two interesting features: the energy separation between flat bands is mostly even, and the location of the projected states alternates between the top and bottom layers. For instance, we find electronic states near $-2.68$~eV at the blue $\mathbf{M}$ from the top layer,  it follows by states projected in the bottom layer (red) at $-2.58$~eV, then blue ones at $-2.48$~eV, and then consecutively. In Figs.~\ref{Fig4} (c), (d), and (e), we show top and side views of the real space along with corresponding wavefunction norms, $|\Psi|$, for the indicated electronic states in Fig.~\ref{Fig4}(b). Here, we observe that while wavefunctions are delocalized in the armchair directions, they are finite along the red line that corresponds to the zig-zag direction. Additionally, the side view shows that $|\Psi|$ are in both G layers, indicating their hybridization that allows the observed resonant behavior.

These discrete states have an energy difference of about $100$~meV or smaller, and are located alternately at the top and bottom layers. On the red ${\mathbf{M}}$ point, an equal behavior is found but changing red by blue and vice versa. Between both $\mathbf{M}$ points, an overlapping between projected states from both layers, resulting in an electronic hybridization or strong coupling between layers due to vdW interactions, is found. This latter is consistent with recent experimental observations in 1D moir\'e superlattices from double-wall carbon nanotubes,  where a strong intertube coupling effect was measured.\cite{1dmoire} It has been shown that this coupling is important for engineering the electronic structure and optical properties of moir\'e superlattices.\cite{1dmoire2} 

\begin{figure}[h]
  \includegraphics[width=0.50\textwidth]{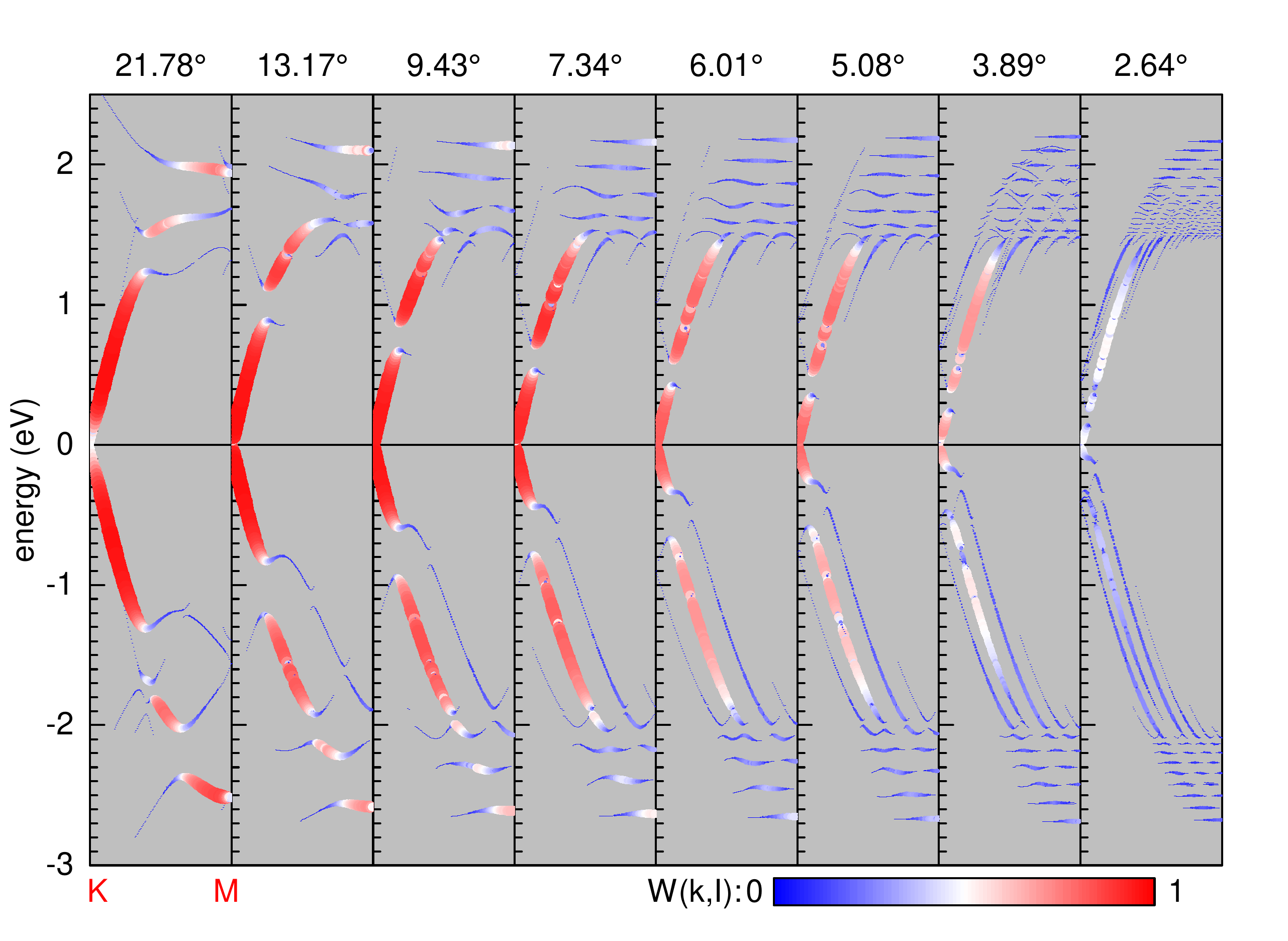}
  \caption{EEB in $\mathbf{K}-\mathbf{M}$ for TBLG with different
    angles. $W(k,I)$ is proportional to dots size and
    color. $E_F=0$~eV.}
  \label{Fig5}
\end{figure}

Finally, we discuss the relevance of $\theta$ as a new parameter for band engineering in GMSs.  Upon the unfolding approach, we find new band gaps that appear below $-2.0$~eV and above $+1.5$~eV, which are equally-spaced in energy near $\mathbf{K}$. Fig.~\ref{Fig5} shows EEBs along the $\mathbf{K} \to
\mathbf{M}$ trajectory for different TBLGs, where discrete electronic states arise. Recall that electronic properties change drastically along this trajectory with energies around $\pm 2.5$~eV, involving $V_{pp\sigma}$ interactions between $2p_{z}$ orbitals from both G layers. As $\theta$ decreases, these
gaps turn into a discrete or localized electronic states evenly-spaced, resembling G quantum dots. By varying $\theta$, the number and energy difference of these evenly-spaced localized electronic states can be tuned. As the energy separation between discrete states become smaller than $100$~meV when $\theta \to 0$, making possible electron-phonon driven processes. These latter might been of interest in ultrafast absorption and emission process in the UV region,\cite{Novko2019} as well as to explain superconductivity at very small angles. \cite{Wu2018,Lian2019} We would like to stress that this is the
primary theoretical evidence using DFT, that a set of flat bands emerging into discrete states because of the interaction between G layers, which can be tuned with the interlayer angle.

\section{Conclusions}

The angle-dependent evolution of graphene moir\'e superlattices electronic properties is discussed within the density functional theory by unfolding supercell bands onto the graphene primitive cell. The unfolding reveals electronic changes upon a second graphene sheet due to the interaction between out-of-plane orbitals ($2p_z$), forming different kinds of gaps and a band splitting. The first kind of gaps are closed to $\mathbf{K}$ and approach the Fermi level as the interlayer angle decreases. We confirm that the origin of these gaps is because of Dirac cones interlayer interactions. The second kind is given by many gaps close to $\mathbf{M}$, where the flat band electronic degeneration is broken due to $V_{pp\sigma}$ interactions of the out-of-plane orbitals. When the interlayer angle goes to zero, theses gaps turn into discrete-like states, which are evenly-spaced in energy with gaps around  100~meV and smaller, and increase in number. These gaps are in the order of the phonon energies of Graphene, where the $E_{2g}$ is at an energy ca. $190$~meV. A strong interlayer coupling gives rise to such gaps, and thus, they depend on the interlayer angle. Finally, at low energies in the valence and along the $\mathbf{\Gamma}-\mathbf{K}$ and $\mathbf{\Gamma}-\mathbf{M}$ trajectories, the band divides into two, showing a hybridization between orbitals from different layers. The results discussed here demonstrate direct evidence of the electronic discretization into a highly-anisotropic and well-ordered states of graphene moir\'e superlattices, which can be relevant  to explain electron-phonon assisted effects recently observed in experiments.

\begin{acknowledgements}
Authors acknowledge partial support from DGAPA-UNAM grant PAPIIT~IN109618 and CONACYT grants A1-S-14407 and 1564464. FSO also acknowledges IPICYT National Supercomputing Center  grant TKII-2020-FSO01. 
\end{acknowledgements}

\appendix

\section{Uncovering the interaction in TBLG}\label{graph}

\begin{figure}[h]
  \includegraphics[width=0.45\textwidth]{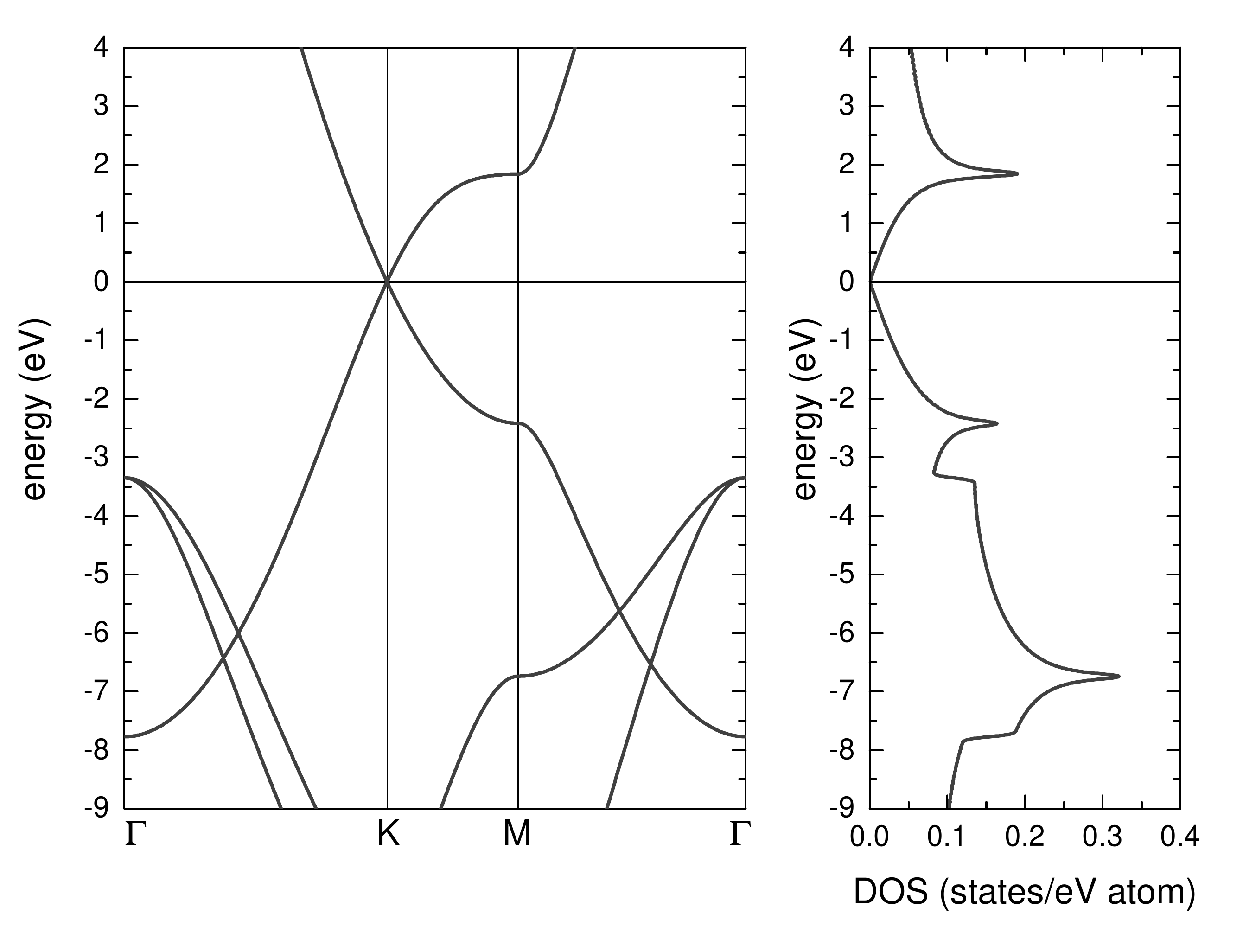}
  \caption{Electronic bands of pristine G, as well as DOS obtained using DFT. Here $E_F=0$~eV.}
  \label{S1}
\end{figure}

\begin{figure}[h]
(a)\includegraphics[width=0.45\textwidth]{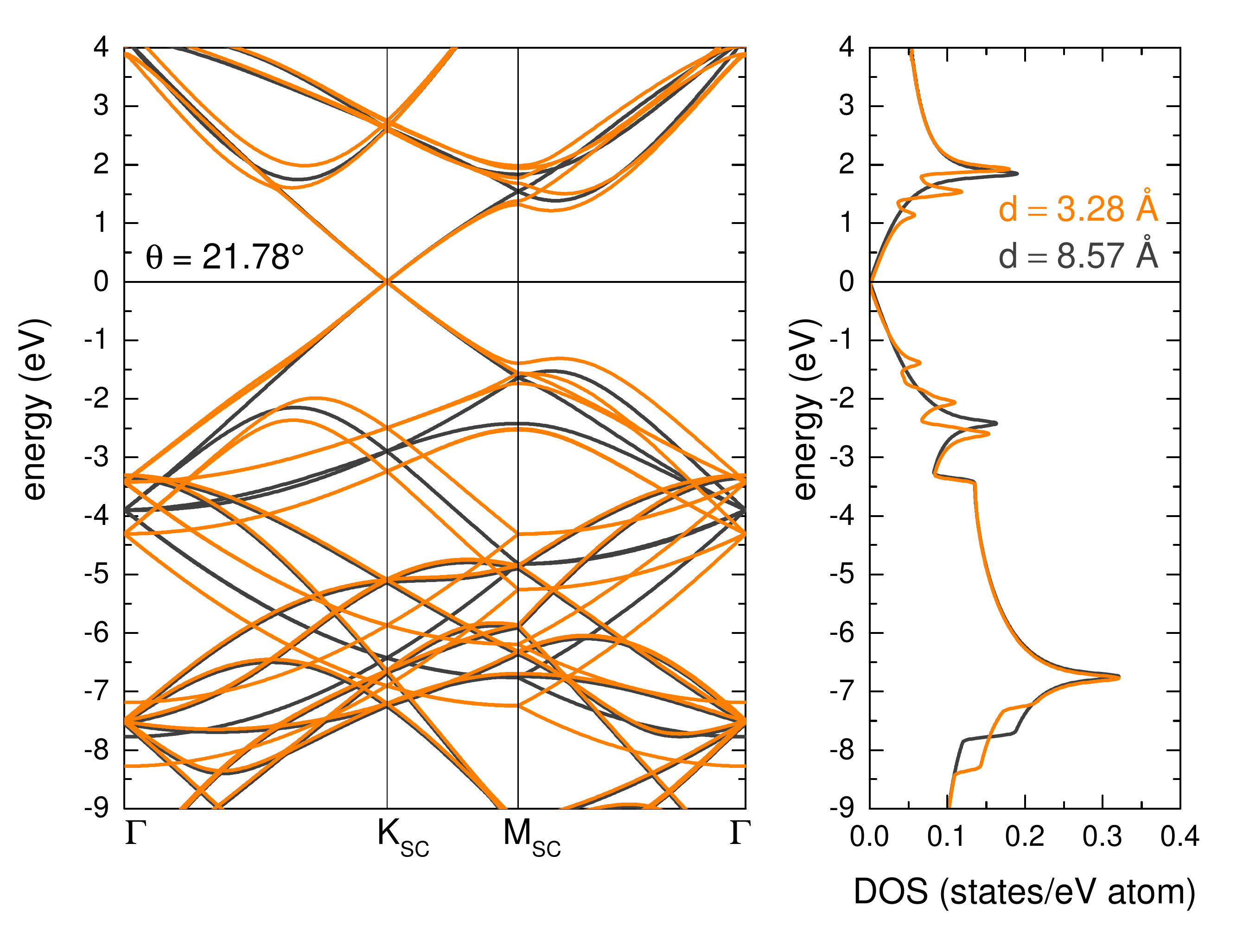}
(b)\includegraphics[width=0.45\textwidth]{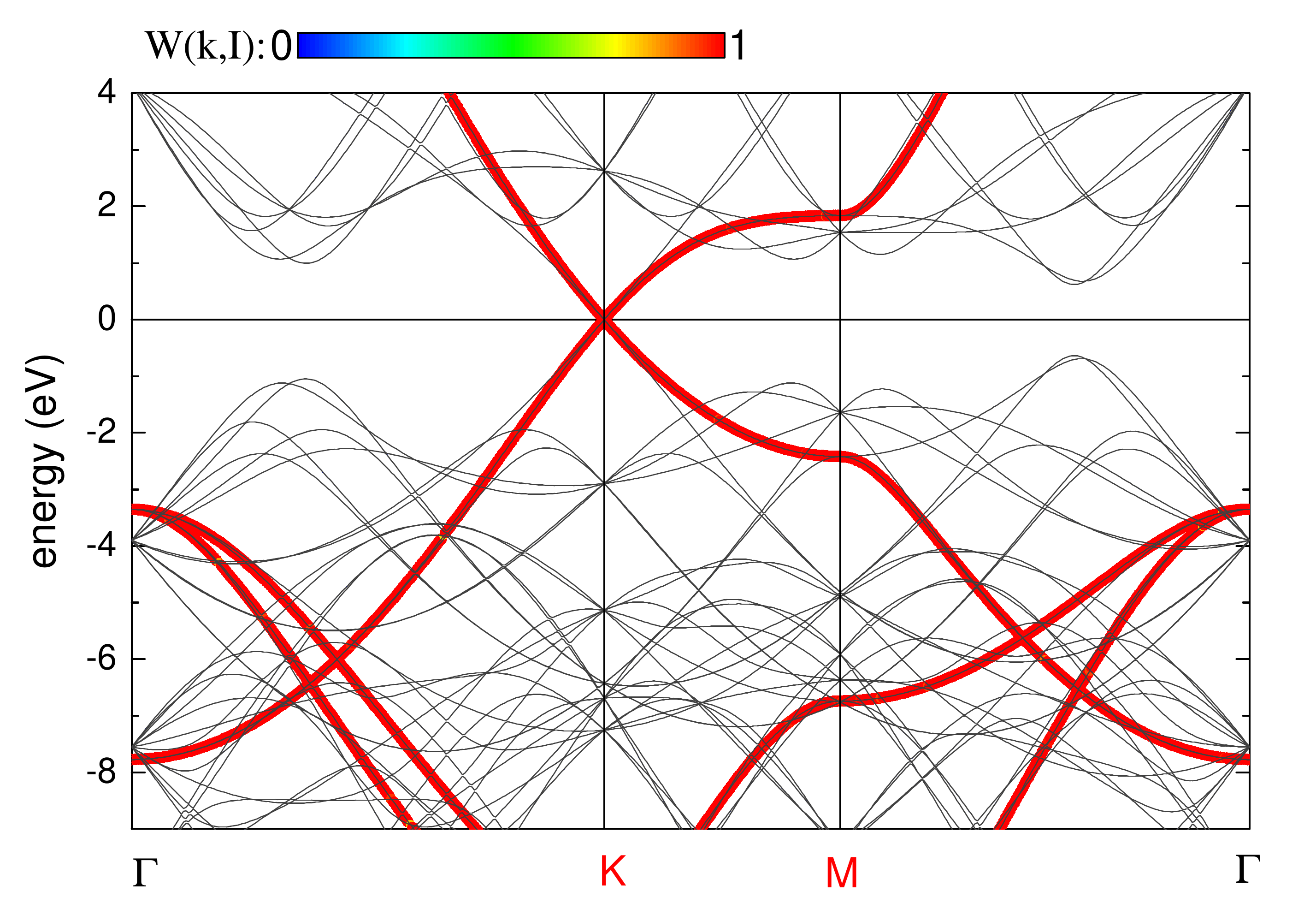}
(c)\includegraphics[width=0.45\textwidth]{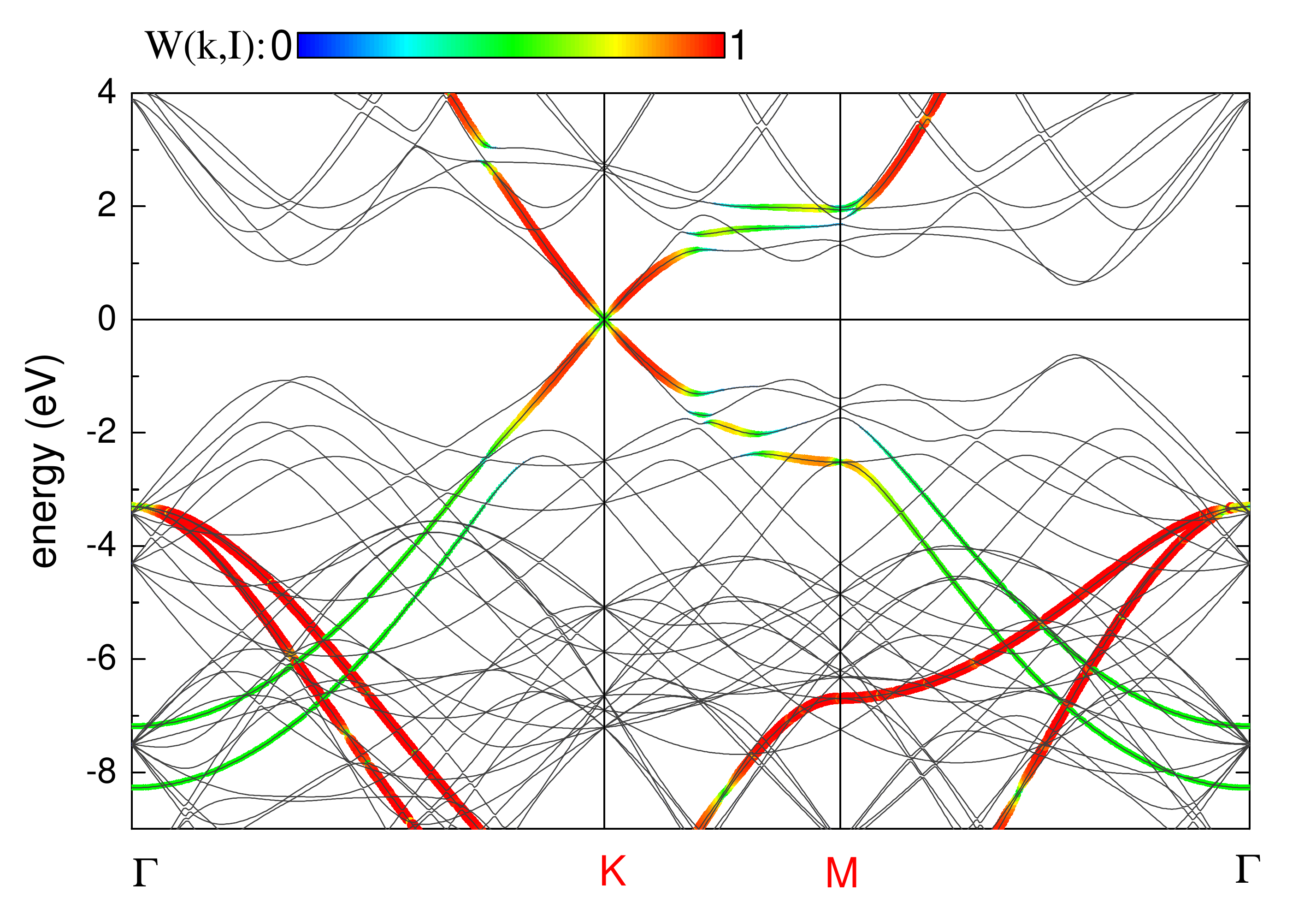}
  \caption{(a) Electronic structure and DOS of \textsc{sc} TBLG (2,1) and $\theta=21.78$ at a separation distance of $d=3.28$~\AA, in orange and at $d=8.57$\AA, in black. (b) EEB of TBLG (2,1)  at $d=8.57$\AA. (c) EEB of TBLG (2,1)  at $d=3.28$\AA. $W(k,I)$ is proportional to dots size and color, and $E_F=0$~eV.}
  \label{S2}
\end{figure}

The detailed procedure for unveiling the interaction in TBLG using the unfolding approach is described next. Let us consider the pristine G as the reference system. The DFT calculated electronic structure and DOS of
the G primitive cell are in Fig.~\ref{S1}. Suppose now that a second layer is placed \emph{on top} of the first, but let us neglect the layers interaction. It can be done in a DFT calculation by placing the second layer at a distance far enough to nullify the interaction. As described in section \ref{TBGL}, under commensurate considerations, the system will form a superlattice \textsc{sc} characterized with indices $m$ and $n$. The high-symmetry points are $\mathbf{K}_\textsc{sc}$, $\mathbf{M}_\textsc{sc}$, and $\mathbf{\Gamma}$, where we have dropped the subscript for the reciprocal lattice origin since it is always the same. The band structure and density of states for a non-interacting system with $(m=2,n=1)$ is shown as the black curve of Fig. \ref{S2}(a). 

At the non-interacting system, the band structure going through a path that connects the high-symmetry points $\mathbf{\Gamma} \to {\mathbf{K}_\textsc{sc}} \to {\mathbf{M}_\textsc{sc}} \to \mathbf{\Gamma}$, is shown as black lines in Fig.~\ref{S2}(a). Upon projection of the eigenstates of the \textsc{sc} onto the orbitals of our reference G (see Eq.~2 of Ref.~\citenum{vdw11}), which can be done in a localized basis set as in this case, we obtain the spectral weights $W(\mathbf{k},I)$. For the non-interacting case, the spectral weights are either 1 or 0, meaning that \textsc{sc} wavefunctions  completely map onto the G primitive cell wavefunctions. In contrast, when considering the equilibrium interlayer distance, where the layers interact, the band structure along the \textsc{sc} path (Fig.~\ref{S2}(a) orange curve), shows significant deviations from G discussed in the manuscript. The difference between the black and orange lines goes beyond the high-symmetry points, where the Bragg diffractions occur. Furthermore, the spectral weights along the $\mathbf{\Gamma} \to {\mathbf{K}_\textsc{sc}} \to {\mathbf{M}_\textsc{sc}} \to \mathbf{\Gamma}$ path have a full distribution of values, meaning that the mapping of the \textsc{sc} onto the G system is no longer perfect, and the interlayer interaction caused it. 

\begin{table}[h]
   \caption{$E_{\text{VHS}}$ of the first gap below the Fermi level, as well as the energy difference between occupied and empty VHS close to Fermi level, $\Delta E_{\text{VHS}}$. TBLG normalized Fermi velocity with pristine G, $\tilde{v}_\text{F} = v_F\text{(TBLG)}/v_F\text{(G)}$. }
  \label{table2}
  \begin{ruledtabular}
  \begin{tabular}{crccccrrr}
    $(m, n)$ & $\theta (^{\circ})$ & $E_{\text{VHS}} (eV)$ & $\Delta E_{\text{VHS}}$ (eV)& $\tilde{v}_\text{F}$ \\ \hline 
    (2,1) & 21.78    & -1.30 & 2.53 & 0.97\\
    (4,2) & 21.78    & -1.30 & 2.54 & 0.97 \\
    (5,3) & 16.42    & -1.02 & 2.01 & 0.96 \\
    (3,2) & 13.17    & -0.82 & 1.66 & 0.95 \\
    (6,4) & 13.17    & -0.82 & 1.68 &  0.95 \\
    (7,5) & 10.99    & -0.70 & 1.41 & 0.94 \\ 
    (4,3) & 9.43      & -0.59 & 1.19 & 0.92 \\
    (5,4) & 7.34      & -0.44 & 0.89 & 0.88 \\
    (6,5) & 6.01      & -0.34 & 0.70 & 0.84 \\
    (7,6) & 5.09      & -0.26 & 0.54 & 0.79 \\
    (9,8) & 3.89      & -0.17 & 0.35 & 0.67 \\
  (11,10) & 3.15    & -0.12 & 0.26 & 0.58 \\  
  (13,12) & 2.64    & -0.09 & 0.21 & 0.52
  \end{tabular}
 \end{ruledtabular}
\end{table}

Among the effects that are easy to identify without the need of the unfolding are $E_{\text{VHS}}$, the energies of the van Hove singularity below the Fermi level, $\Delta E_{\text{VHS}}$, the energy difference between van Hove singularities peaks closer to the Fermi energy, as well as the velocity ratio, $\tilde{v}_\text{F} = v_F\text{(TBLG)}/v_F\text{(G)}$, of the charge carriers in the valence band. These charge carriers velocities are calculated from the fitted slopes, as shown in Fig.~\ref{S3}(c). These values are in Table~\ref{table2}, which compare well with experimental results based on ARPES measurements in Ref.~\citenum{yin_selectively_2016}. From here, we estimate a value of $E_{\text{VHS}} =1.6$~eV,  for an angle of ca. $\theta=21^\circ$. Our value for $\theta=21.78^\circ$ is $1.3$~eV, that is, indeed, a discrepancy of around $18\%$. This difference is anticipated when utilizing DFT, which underestimates the Fermi velocity of graphene by about $15\%$ because of the lack of many-body effects, as shown by GW calculations. \cite{PhysRevLett.101.226405} Also in Table~\ref{table2}, the $\Delta E_{\text{VHS}}$ values are displayed, which were accessed through optical measurements, \cite{moon2013optical} which are also in good agreement with our results. In both cases, differences are expected because of both DFT and measurement accuracies, including the experimental estimation of $\theta$.

\section{Gaps close to the Dirac cones}\label{dirac}

\begin{figure}[h]
  \includegraphics[width=0.48\textwidth]{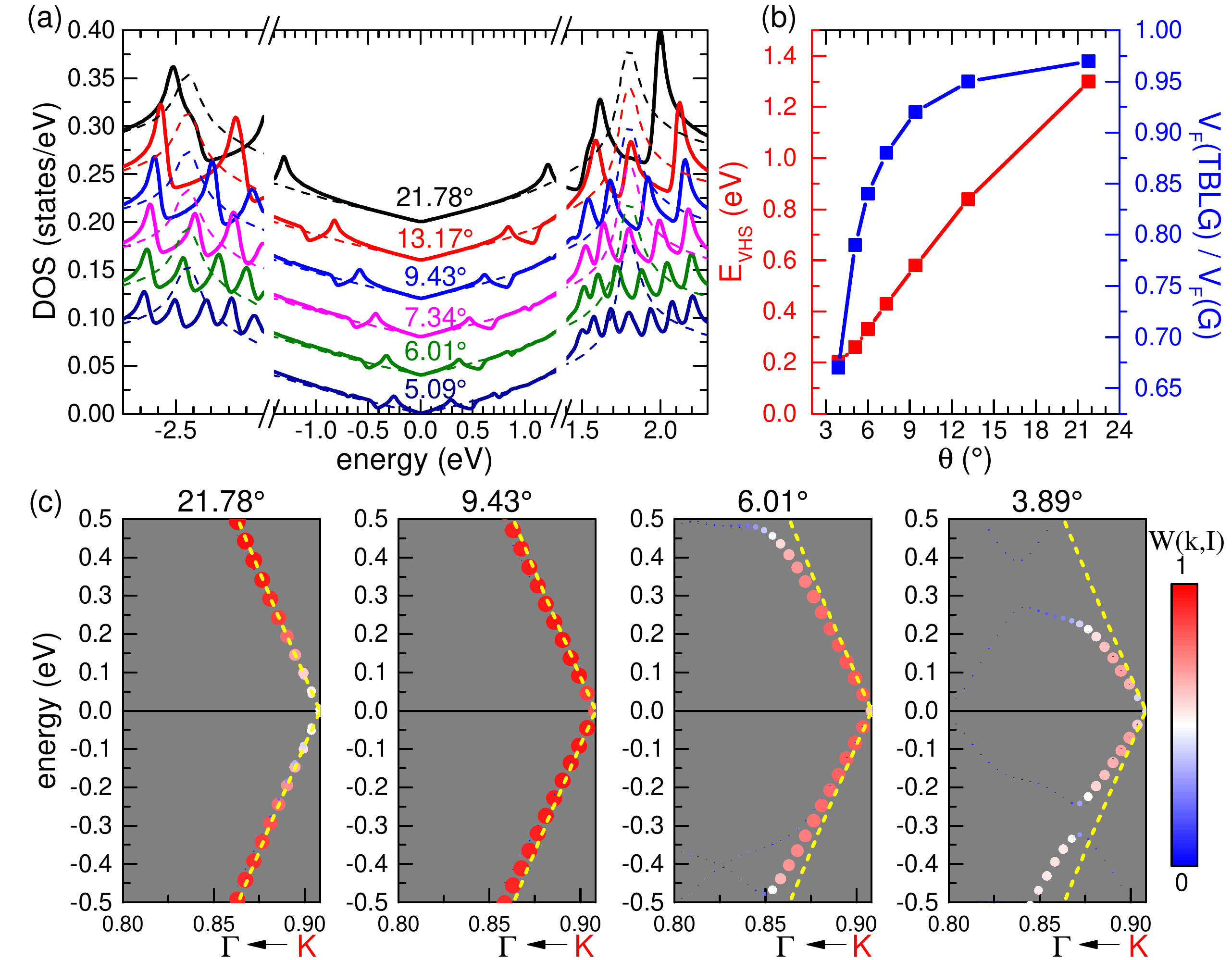}
  \caption{(a) DOS per atom close to $E_F$ as a function of $\theta$. Plots are vertically shifted, and dashed lines correspond to pristine G, included for clarity. (b) $E_{\text{VHS}}$ of the first gap below $E_F$, and normalized Fermi velocity, $\tilde{v}_\text{F} = v_F\text{(TBLG)}/v_F\text{(G)}$, as a function of $\theta$. (c) EEB close $E_F$ and around $\mathbf{K}$ for TBLG with different $\theta$.  $W(k,I)$ is proportional to dots size and color. For comparison, we plot the linear dependence of pristine G band (dashed-yellow line). }
  \label{S3}
\end{figure}

Here, we focus on studying the gaps discussed above, and their corresponding VHS as a function of $\theta$, by inspecting DOS of different TBLG systems in Fig.~\ref{S3}(a). For comparison, we also include DOS of pristine G. The electronic gaps are identified by the depletion in DOS (solid lines) compared with pristine G (dashed lines). For $\theta = 21.78^{\circ}$, the first empty and occupied VHS are at about $+1.2$ and $-1.3$~eV, respectively, with an energy difference, $\Delta E_{\text{VHS}}$ of about $2.5$~eV . A systematic decrease of $\Delta E_{\text{VHS}}$ is found, as $\theta$ also does. Before, we mentioned that
these two gaps which originate the DOS peaks are close to $\mathbf{K}$. Moreover, the $\pi-$band dispersion depends on the relative angle between layers, inducing a variation of the charge carriers group velocity. We particularly find a reduction of the group velocity along the $ \mathbf{\Gamma}-\mathbf{K}$ direction, as $\theta \to 0^\circ$. Fig.~\ref{S3}(b) shows EEB for TBLG with $\theta = 21.78, \, 9.43, \,6.01,$ and $3.89^{\circ}$ for $k-$points near $\mathbf{K}$ approaching from $\mathbf{\Gamma}$. Notice that spectral weights also change as a function of $ \theta $, becoming smaller when $\theta \to 0^\circ$, meaning that the interlayer interaction is stronger at small angles. The slope of the linear dispersion
adjacent to $\mathbf{K}$ changes, meaning a reduction of the Fermi velocity ($v_{F}$) of the charge carriers. This reduction has been previously reported \cite{uchida2014atomic,brihuega2012unraveling,trambly2010localization}. For the largest $\theta$, the slope change is minimal compared to pristine G. So far, we have shown that the electronic properties in TBLG depend on the interlayer interaction of $2p_{z}$ orbitals, which can be modulated with the angle between layers. The coupling is more significant as the moir\'e supercells area is large. These results are summarized in Table~\ref{table2} and Fig.~\ref{S3}(b), where it is clearly shown that $\Delta E_{11}$ and $v_{F}$ tends to zero as $\theta \to 0^{\circ}$. This latter leads to the localization of Dirac electrons in regions with AA-stacking in real space \cite{trambly2010localization}, while at such energies TBLGs behave more likely as pristine G layers when $\theta\to 30^{\circ}$, as the interaction is small. These results are in agreement with other models. \cite{de2012numerical,dos2012continuum}

\section{Tight-binding model}\label{tb}

The tight-binding model used here is formulated within the Slater-Koster approach, considering only $p_z$ orbitals. For both in-plane and out-of-plane interactions, we assumed an exponential distance dependence, $\exp{-\alpha(\frac{d}{d0} -1 )}$, where $d$ is the distance between different atoms, $d_0^{\text{in}}=1.42$~\r{A} is the first-neighbors distance between in-plane atoms. In comparison, the first-neighbors distance between out-of-plane atoms is $d_0^{\text{out}}=3.35$~\r{A}.  Interactions type $\sigma$ and $\pi$ between $p_z$ orbitals are considered for both sublattices and among them. The parameters were fitted to Hamiltonian elements after the DFT SCF Hamiltonian ''Wannierization'' \cite{Pizzi_2020}, being thus, an
orthogonal model. The in-plane hopping parameters are:
\[V_{pp\pi}^{AA} =0.8316\,\text{eV}, \,\,\,\, V_{pp\pi}^{AB} = -2.89\,\text{eV}, \,\,\,\, V_{pp\sigma}^{AB} = -2.89\,\text{eV},\] 
\[\alpha^{AA}=1.70, \quad \alpha^{AB}=2.60, \quad \text{and} \quad d_0^{\text{in}}=1.42~\text{\r{A}}.\] 
Note that in-plane interactions type $V_{pp\sigma}$ is always null
since the cosine director is zero.  The out-of-plane parameters are:
\[V_{pp\sigma}^{AA'} =0.299\,\,\text{eV}, \quad V_{pp\pi}^{AA'} = -1.035\,\,\text{eV}, \]

\[\quad V_{pp\sigma}^{AB'} =0.268\,\,\text{eV}, \quad V_{pp\pi}^{AB'} = -0.819\,\,\text{eV}, \]  
\[\text{here,} \quad\alpha^{AA'}=\alpha^{AB'}= 8.1, \quad \text{and} \quad d_0^{out}=3.35~\text{\r{A}}.\]

\begin{figure}
  \includegraphics[width=0.5\textwidth]{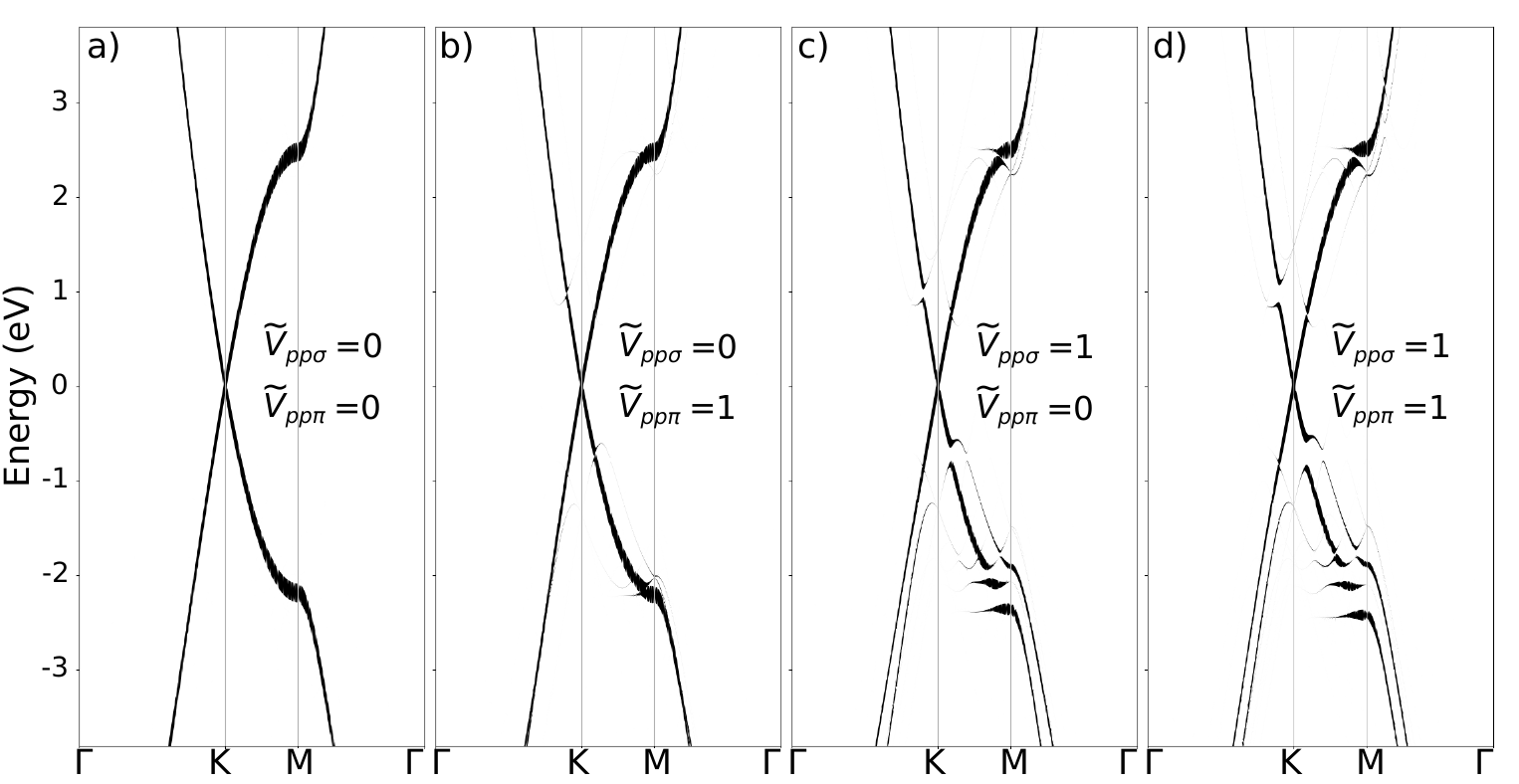}
  \caption{Effect of different interlayer interaction parameters based on the tight-binding model for TBLG (4,3) with $\theta=9.43^\circ$.
The reduced parameters are defined in terms of the fitted values (starred): $\tilde{V}_{pp\sigma}=V_{pp\sigma}/V^*_{pp\sigma}$; $\tilde{V}_{pp\pi}=V_{pp\pi}/V^*_{pp\pi}$}
  \label{S4}
\end{figure}

\bibliography{angle}

\end{document}